\documentclass[runningheads,a4paper]{llncs}

\usepackage{amssymb}
\usepackage{graphicx}
\usepackage{cite}
\usepackage{url}
\usepackage[affil-it]{authblk}

\urldef{\mailsa}\path|{csummers@gracenote.com, ppopp@gracenote.com}|  

\begin{document}

\mainmatter  

\title{Large Scale Discovery of Seasonal Music\\
	From User Data}

\titlerunning{Large Scale Discovery of Seasonal Music}

\author{Cameron Summers \and Phillip Popp}

\authorrunning{Cameron Summers\and Phillip Popp}

\institute{Gracenote\\
Emeryville, CA United States\\
\mailsa\\}

\toctitle{Lecture Notes in Computer Science}
\tocauthor{Authors' Instructions}
\maketitle

\begin{abstract}
The consumption history of online media content such as music and video offers a rich source of data from which to mine information. Trends in this data are of particular interest because they reflect user preferences as well as associated cultural contexts that can be exploited in systems such as recommendation or search. This paper classifies songs as seasonal using a large, real-world dataset of user listening data. Results show strong performance of classification of Christmas music with Gaussian Mixture Models.

\end{abstract}

\section{Introduction}

Consumption of media content such as music and video often exhibits seasonal patterns. Identifying and understanding these seasonal contexts can improve the quality of recommendations as shown by \cite{shin} and provide useful explanations for the recommendations that are made, improving the user experience \cite{wang07}. The cultural context of the season often extends to other domains beyond music listening, linking music recommendation with other recommendations systems. The importance of context in music can be readily observed in industry where flags for seasons such as Christmas are often used \cite{spotify_xmas}. However, the task of manually labeling specific content as connected to a season is challenging because these connections have a distributed nature - varying by geographic region, language, and time - and expert curation is time intensive and costly. We investigate the feasibility of labeling seasonal content by classification with user listening data.

Previous research has studied the dynamics and classification of time series signals. In the web search domain, \cite{kulkarni} showed that queries could be classified by their change in popularity over time using features in the signal. \cite{shokouhi11} classified seasonal web search queries using Holt-Winters decomposition on a small data set to improve time-sensitivity in search results. In music listening signals, \cite{park}, \cite{carneiro}, and \cite{hidasi} show how analysis of temporal dynamics of music listening are useful for recommendations systems and look specifically at seasonality. However, to our knowledge there is no published work that attempts to exploit the temporal analysis of music listening data for automated labeling of seasonal music content.

\section{Methods and Materials}

\subsection{Approach} \label{approach}

Listen counts of a track will peak at a specific period of time if it has an association with that period, such as a Christmas track on December 25th. This pattern can be exploited by training a classifier with features of this signal. The features used in this paper are daily listen counts of a track for a window of time localized around the target season. To control for the significant differences in the overall popularity of tracks in a large data set, we normalize the listen counts of each track across the selected periods. The listening rates, $R$, are described in Equation \ref{eq:feature}.

\begin{equation}
  R_{ij} = \frac{ \sum_{k=1}^{u} c_{ijk} } { \sum_{l=1}^{w} \sum_{k=1}^{u} c_{ilk} } ,  j \in w  \; .
  \label{eq:feature}
\end{equation}

where $ c_{ijk} $ is the number of listens by the k\textsuperscript{th} user in the j\textsuperscript{th} period of time for the i\textsuperscript{th} track, $c_{ilk}$ is the number of listens by the k\textsuperscript{th} user in the l\textsuperscript{th} period of time for the i\textsuperscript{th} track, $w$ is a set of discrete of periods of time, and $u$ is the number of users. 

For classification, we chose the Gaussian Mixture Model (GMM) with full covariance matrix because it is fast to train and the listening rates resemble a normal distribution. A GMM is trained using tracks from the target season in a training portion of the data set, and classification is performed on the test set. 

\subsection{Dataset} \label{ds}

\begin{center}
\begin{tabular}{ p{4cm} p{6cm} }
\hline
  Number of Records & 4,819,992,847\\
   Number of Users  & 1,648,796\\
  Number of Tracks & 13,227,376\\
  Date Range & January 2012 - February 2013\\
\hline
\label{table:dataset_stats}
\end{tabular}
\end{center}

This study uses an internal Gracenote dataset of online radio listening records in North America with some basic statistics of the dataset shown in Table \ref{table:dataset_stats}. Each record of the dataset represents one listen of a track by one user and provides User ID, Date, Time, and Track ID. From the Track ID some associated metadata such as track name and album name is used for keyword search and post-experiment analysis. It is necessary to use a large dataset to get good classification results as shown in section \ref{xmas_exp}. Other public datasets similar to \ref{ds} such as "Last.fm Dataset - 1K users" dataset available at \url{http://www.dtic.upf.edu/~ocelma/MusicRecommendationDataset/lastfm-1K.html} are too small.

\subsection{Experiment - Christmas} \label{xmas_exp}

We chose Christmas as the target for seasonal music identification because of its popularity and large volume of associated music. We hypothesize that a classifier trained with features in section \ref{approach} can identify Christmas tracks. We generated an initial ground truth of Christmas tracks by searching for "Christmas" keyword in the track name and album name - totaling 87,554 Christmas tracks or 0.7\% of the entire track population - and maintained a second list of tracks without the keyword. This is not a comprehensive list of Christmas tracks, but this method provides a relatively clean ground truth. Expert curation of a ground truth is infeasible with such a large dataset, and using tags from external sources is error prone.
 
\begin{figure}[!ht]
\centerline{\includegraphics[height=10cm]{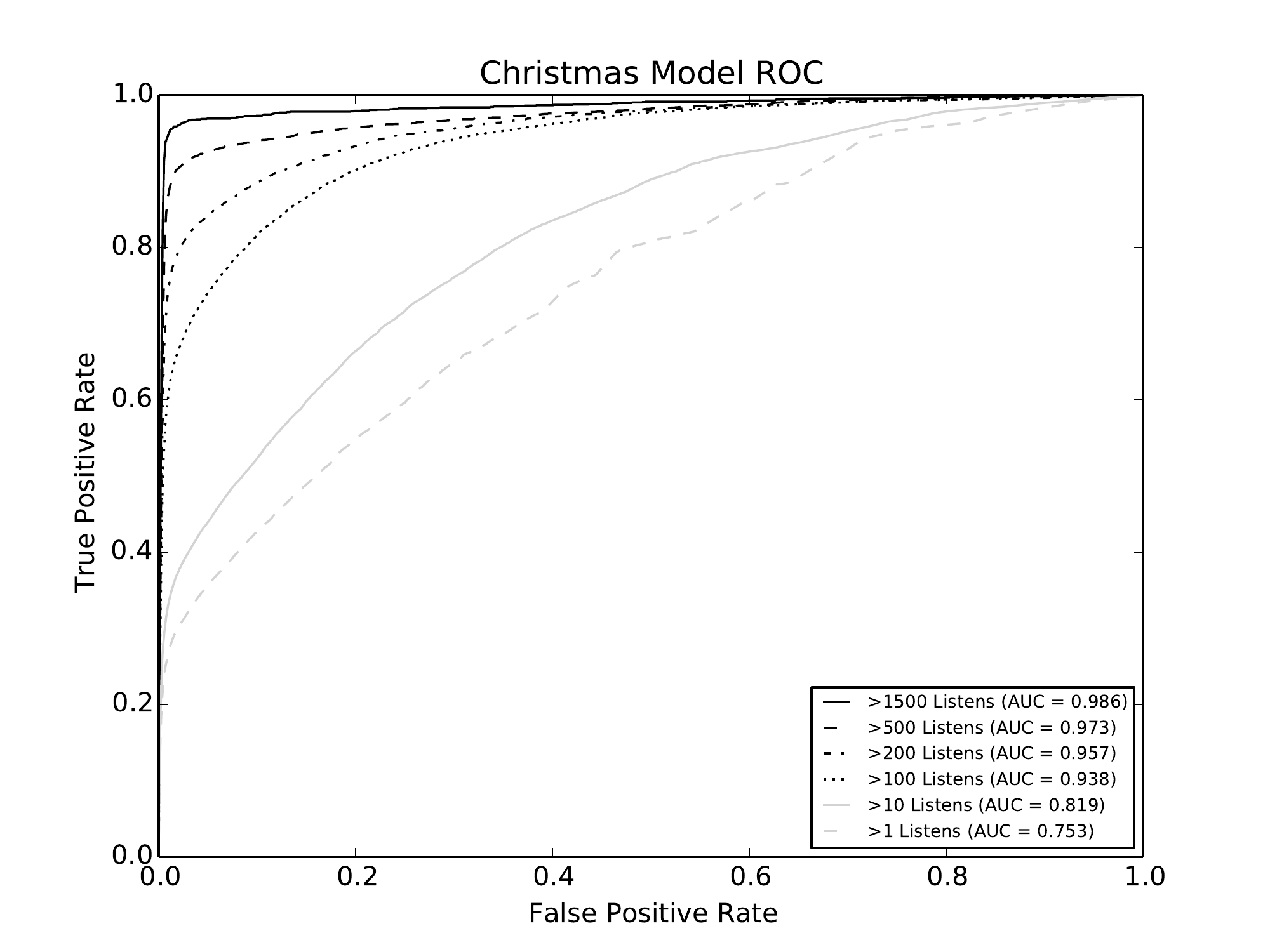}}
\caption{Test set ROC and AUC of the Christmas classifier on subsets of the dataset thresholded by minimum total listens. The 1,500 listen threshold has 335,678 total tracks with 4,732 Christmas tracks.}
\label{fig:xmas_results_roc}
\end{figure}

We chose a consecutive 15 day span centered on December 25th, Christmas, as the listening rate inputs to the classifier. Training and classification (60\% train, 40\% test) using Gaussian Mixture Models were performed on subsets of the dataset given by tracks with more than some minimum total listens in the whole dataset. To validate performance of the Christmas model, ROC and AUC score were calculated on the test set and are in Figure \ref{fig:xmas_results_roc}.

%

\section{Discussion}  \label{discussion}

The performance of the model is quite good even though the ground truth has an incomplete list of Christmas tracks. At the highest threshold, an inspection of tracks with high probability according to the Christmas model without the "Christmas" keyword shows that many are other Christmas songs well-known in North America such as "The First Noel" and "Santa Claus Is Coming To Town." This suggest that performance would likely increase with a more complete list of Christmas tracks.

One notable observation is the change in AUC as the threshold for total minimum listens of track is lowered. Classification suffers when including unpopular tracks. This is likely due to the natural variance in the listen counts of tracks with fewer listens. Normalizing smaller listen counts has a disproportionate effect on computation of listen rates.

The model trained with Christmas tracks could be used to identify other seasonal tracks at different times of the year. One possible application of this would be an "always on" seasonal radio station. This is a topic of future work.

\section{Conclusion}

This study demonstrated on a large, real-world dataset that user listening data could be utilized to detect seasonal music content for Christmas. Classification with a Gaussian Mixture Model showed that the listen rates are sensitive to variance in unpopular tracks and quality results require detection to be performed on a large database of listening records.

\end{document}